# Rental Housing Spot Markets: How Online Information Exchanges Can Supplement Transacted-Rents Data

Geoff Boeing[1], Jake Wegmann[2], Junfeng Jiao[2]

**Abstract:** Traditional US rental housing data sources such as the American Community Survey and the American Housing Survey report on the transacted market—what existing renters pay each month. They do not explicitly tell us about the spot market—i.e., the asking rents that current homeseekers must pay to acquire housing—though they are routinely used as a proxy. This study compares governmental data to millions of contemporaneous rental listings and finds that asking rents diverge substantially from these most recent estimates. Conventional housing data understate current market conditions and affordability challenges, especially in cities with tight and expensive rental markets.[†]

## 1. Introduction

How much does it cost to rent a typical apartment in your city? Answering this basic housing question can be surprisingly difficult. Consider the case of San Francisco in early 2018, one of the US's most expensive cities and a critical site for scholars and practitioners working on affordability, gentrification, residential mobility, and housing justice. According to the most recent (2016) American Community Survey (ACS), a typical two-bedroom unit in San Francisco rented for $2,017/month. We can compare this figure to the most recent (2015) American Housing Survey (AHS)—whose data are only available for the entire San Francisco metropolitan statistical area (MSA)—to find a median two-bedroom rent of $1,722.

In reality, anyone attempting to move to San Francisco or its suburbs in 2018 would have been hard-pressed to find a two-bedroom apartment on the market at these rent levels.

---

The US Department of Housing and Urban Development's (HUD) contemporaneous Fair Market Rent (FMR)—published to determine Section 8 voucher payments to landlords—was $3,121 for a 40th-percentile two-bedroom unit in San Francisco's MSA. This figure comes closer to what someone moving to San Francisco may have had to pay but is still possibly too low. Data from Zillow (a for-profit company that aggregates rental listings) estimated that the median two-bedroom unit in the city of San Francisco rented for $4,197—double the ACS estimate (Reese 2018).

How might housing practitioners and scholars make sense of the drastic discrepancies between these estimates? These data sources obviously differ in recency due to varying publication lags, in geography as some cover the core city while others cover the entire MSA, and possibly in sampling biases or respondent errors. However, when considering how much it costs to rent a home, an equally important distinction exists between data that report rents for the *transacted market* (existing leases) and those that report on the *spot market* (current asking rents on the open market). That is, does the data product tell us what a typical existing renter pays each month, or rather what someone must pay if they currently seek housing? Transacted market data could diverge significantly from the spot market due to publication lags, market lags, subsidies, rent control, and length-of-stay discounts.

Most readily-available sources of rents in US communities report data for the transacted market rather than the spot market. While such data sources are valuable, they do not tell us what someone currently must pay to acquire housing. This is a critical gap. According to the 2016 ACS, 37% of US renters reported having moved since the beginning of 2015. Yet much of this recent market activity remains illegible to scholars, planners, and advocates, inhibiting important research and policy development on pressing issues like the neighborhood-level effects of market-rate housing production on displacement, the impact of new transit stations on rental affordability, and the functioning of housing markets following natural disasters. In contrast, studies of owner-occupied housing do not suffer from this data shortage—price information in both transacted and spot markets is widely available. Yet rents are often more responsive than purchase prices to market trends and exogenous shocks. These information asymmetries constrain the questions that researchers can answer and even ask about housing markets.

How well do readily-available rent estimates represent the rental spot market in different cities? This study analyzes Craigslist rental listings to investigate the legibility of asking rents and the inconsistent stories that existing data sources tell us about current affordability conditions in tight markets. We find that existing sources typically understate how expensive it is for people to relocate, with particular ramifications for intermetropolitan movers and voucher holders. First, we discuss currently-available rental data sources and their trade-offs, then summarize existing research on the economic implications of rapidly rising rents. Next, we describe our data sets and methods to analyze the discrepancies between existing data and the realities of the spot market. To do so, we compare the rental markets of 25 MSAs using ACS, AHS, FMR, and Craigslist data. Finally, we discuss implications for planners and suggest how online rental listings and other sources of spot market data could be made more widely available to close this critical information gap.

## 2. The Rental Housing Data Landscape

### 2.1. Rental Housing Affordability

Few planning topics generate as much public concern today as rental housing. Recent decades have witnessed an across-the-board decline in rental affordability (Myers and Park 2019), particularly for low-income renters (Chan and Jush 2017). This trend most famously impacts the US's large coastal cities with robust economies and stiff housing supply constraints (Sassen 2001; Saiz 2010; Glaeser et al. 2005). But today, affordable housing for low-income renters—while never abundant—is growing scarcer even in less-expensive inland metropolitan areas with few natural or regulatory obstacles to development (Immergluck et al. 2017).

Metropolitan inversion—the revalorization of residential and commercial real estate in urban cores vis-à-vis outlying regions—is now occurring in almost all US urban areas following 75 years of centrifugal capital flows (Edlund et al. 2016). In areas where rentals have long constituted a larger share of the housing market—and where economic, cultural, and civic opportunities are growing—access to rental housing is dwindling for underprivileged families, even in otherwise affordable cities like Houston and Phoenix (Pfeiffer and Pearthree 2018). Accompanying this trend of declining rental affordability is a trend of declining home ownership. Citing mechanisms ranging from demographic shifts to persistent income inequality, three of the four forecasts in a recent *Cityscape* issue predict US homeownership rates in 2050 will fall from 2015's 63% rate, which itself was already substantially down from the all-time high of 69% in 2004 (Acolin et al. 2016; Myers and Lee 2016; Nelson 2016; Haurin 2016). As rental housing becomes less affordable, it is simultaneously—and unsurprisingly—becoming more important.

Despite the critical significance of rental housing affordability, US data on rents are inconsistent and incomplete compared to those on owner-occupied housing. Researchers analyzing the latter have two primary information resources: property tax assessor data (roughly covering the transacted market) and multiple listing service (MLS) data (roughly covering the spot market). Neither serves every purpose perfectly. Property tax records are individually searchable online across the US, including in all populous urban and suburban counties, but in some cases tax assessors will not directly provide bulk data to researchers. In such instances, the data can be obtained for a fee from third-party commercial data aggregators. However, tax assessors use widely varying methodologies across counties and states that often diverge from pure market valuations (Martin and Beck 2018). Conversely, MLS data sometimes omit certain transactions or only include listed prices instead of consummated, transacted prices, which themselves are a matter of public record in most states and available in online property tax records. These two data sources complement each other's weaknesses and are essentially universally available.

No equivalent of either data source exists for rents. Accordingly, most housing research relies on real estate prices, even when important research questions revolve around rents (e.g., Freemark 2019). For instance, studies examining whether new rail transit stations raise surrounding property values usually rely on single-family home prices simply because they are obtainable (e.g., Hess and Almeida 2007; Pan 2013; Cao and Lou 2018). However, rents are likely more responsive to transit proximity than single-family home prices are: rental housing tends to contain smaller and non-family households (both likelier to use transit) and rents can respond more rapidly to new local amenities than property sales can. The specter of rising rents displacing transit-dependent low-income families from station areas has alarmed social justice advocates and transportation planners seeking to increase transit ridership (Zuk et al. 2018). Yet real estate prices in station areas receive more research attention than rents because of disparities in spot market data availability and thus study tractability.

**2.2. Rental Housing Data Sources**

Table 1 summarizes the most common rental housing data sources and the trade-offs inherent in choosing between them. These sources fall into four categories. Those in the first, *aggregated federal government sources*, are the most widely known and used. They include the Census Bureau's ACS and AHS as well as HUD's FMR data. These data are aggregated: they summarize rents for groups of housing units within some geographical area. However, full distributions (rather than summary statistics) of rents would offer much more useful information. Although some distributions of rents can be obtained from the aggregated federal government sources, they are limited and often cannot be cross-tabulated with other useful variables (cf. Immergluck et al. 2017).

The second category, *disaggregated federal government sources*, allows the researcher to analyze data from individual housing units. This includes the widely-available AHS microdata and the ACS's Public Use Microdata Sample (PUMS). However, to protect the privacy of survey respondents, their locations are abstracted to large geographical areas: respectively, MSAs of varying size and Public Use Microdata Areas (PUMAs) of approximately 100,000 people each. Some researchers sidestep these limitations by acquiring permission from the Census Bureau to work directly with confidential individual data, but the process to obtain this "Special Sworn Status" is time-consuming and difficult.

The third category, *local government administrative data*, remains more noteworthy for its potential than for its widespread use. Local governments collect these administrative data to meet operational needs and supplement traditional sources, such as the census (Coulton 2008). Molfino et al. (2017) demonstrate the research value of administrative data as an ACS supplement, though they rely on tax assessor data and provide no information about rental housing. Researchers could potentially also use data from rent boards, local bodies charged with administering rent regulations in the small minority of US cities that have them. Autor et al. (2012) use local data to identify formerly rent-controlled properties, but not to assess rents themselves. To access these data, the researchers had to file a Freedom of Information Act (FOIA) request, a nontrivial process.

**Table 1**. Rental housing data sources and their characteristics.

| Category/Examples | Owner | Disagg? | Spatial Coverage | Temporal Coverage | Spatial Resolution | Temporal Resolution | Market Segment | Ease of Access |
|---|---|---|---|---|---|---|---|---|
| *Aggregated fed govt sources* | | | | | | | | |
| ACS data | Federal govt | No | Nationwide | Since 2005 or 2005-2009 | Census block group | Annual | All occupied housing | Widely available |
| AHS data | Federal govt | No | Select MSAs | Since 1973 | MSA | Every two years | All occupied housing | Widely available |
| HUD FMR data | Federal govt | No | Nationwide | Since 1983 | MSA or ZIP code | Annual | Recent movers (unsubsidized housing) | Widely available |
| *Disaggregated fed govt sources* | | | | | | | | |
| ACS PUMS microdata | Federal govt | Yes | Nationwide | Since 2000 | PUMA | Annual | All occupied housing | Widely available |
| AHS microdata | Federal govt | Yes | Select MSAs | Since 1973 | MSA | Every two years | All occupied housing | Widely available |
| Confidential census data | Federal govt | Yes | Nationwide | Nearly unlimited | Housing unit | Annual | All occupied housing | Very difficult |
| *Local govt admin data* | | | | | | | | |
| Rent board admin data | Local govts | Yes | Small number of cities | Various | Housing unit | Quarterly (typical) | Occupied housing on properties subject to local rent regulation | Special permission or FOIA |
| *Proprietary sources* | | | | | | | | |
| Apartment owner survey data | Private companies | Yes/No | Most MSAs | Various | Individual property | Various | Occupied housing on large, unsubsidized properties | Aggregates widely reported; disagg for purchase |
| Proprietary rental listing data | Private companies | Yes/No | Nationwide | Various | Housing unit | Continually updated | Spot market (unsubsidized) | Aggregates rarely reported; disagg unavailable |

The final category comprises *proprietary sources.* These have traditionally taken the form of surveys of apartment property owners (usually limited to some minimum size, e.g., 50 units or more), conducted by market research firms. These firms typically make the resulting data sets available for purchase (at a high price) to researchers, although the news media are often given aggregated synopses. More recently, the rise of Internet-based platforms has allowed for-profit companies such as Zillow, RentJungle, Craigslist, and PadMapper to amass vast proprietary databases of rental listings (Boeing and Waddell 2017; Porter et al. 2019). Their data are tantalizingly rich, detailed, and rare in their ability to describe the spot market in disaggregate form, despite potential sampling biases such as correlations between sociodemographics and preferences to list/search for rental housing online (Boeing 2019). Yet these data remain under the control of private companies that only release information to researchers in limited circumstances.

## 2.3. Advantages and Limitations of Spot Market Data

If made widely available, data sets compiling public rental listings on proprietary web sites would provide an invaluable complement to the constellation of sources listed in Table 1, filling some of the critical spot market gaps that currently exist. Given the pressing need to understand current market conditions and asking rents, many scholars have started collecting samples of publicly-available online rental listings to analyze ad hoc (e.g., Mallach 2010; Halket and di Custoza 2015; Boeing and Waddell 2017; Brown et al. 2017; Im et al. 2017; Schachter and Besbris 2017; Palm 2018).

Much like the governmental data sources, rental listings data have both strengths and weaknesses, as well as both consistencies and inconsistencies with the governmental data. Boeing and Waddell (2017) document these consistencies, including how Craigslist and HUD values correlate across MSAs, but caveat that Craigslist represents advertised rents, not final consummated transactions—an important limitation to remember. Boeing (2019) explores inconsistencies between Craigslist and ACS data, acknowledging the limitations of using ACS rent data but also quantifying Craigslist's sampling biases: in some cities the online listings over-represent whiter, wealthier, and better-educated neighborhoods. In the constellation of imperfect information sources about the rental market, online listings like Craigslist offer the benefits of timeliness and unit characteristics, but have three key limitations: 1) they represent asking—not transacted—rents; 2) they exhibit some sampling biases; 3) they alone do not include demographic data of listers or renters. Nevertheless, spot market data reveal the information landscape and constraints faced by people seeking to change their housing circumstances at a given moment.

As illustrated by the introduction's discussion of San Francisco, spot market rents could differ significantly from transacted rents for four primary reasons: publication lags, market lags, subsidies, and rent control. First, census and other survey data are gathered, compiled, and published with a lag of months or years. Rents and other facts on the ground can change drastically during this delay. Second, due to supply constraints, excess demand, length-of-stay discounts, or even just inflation, the present year's rents could exceed those of past years' leases. Because relocation imposes costs on tenants and landlords, economic theory (Galster 2019) and empirical evidence (Larsen and Sommervoll 2008) suggest that rents for tenants in the transacted market lag behind rents faced by those seeking housing on the spot market. Third, many tenants live in income-restricted dwellings, public housing, or units subsidized with Housing Choice Vouchers (HCV) and Low-Income Housing Tax Credits—substantially impacting the actual rent paid. Fourth and finally, in some high-rent cities—including San Francisco and some of its suburbs—rent control slows the rate at which rents rise for many households (Metcalf 2018). Particularly in such cities—which often anchor rental markets with the most extreme conditions of low vacancy and constrained supply—obtaining information about spot markets rather than rents paid in accordance with previously-transacted leases is critical for shedding light on vexing public policy problems.

One such problem concerns the constraints faced by HCV holders moving within and between MSAs. Substantial evidence suggests that HCV holders, particularly those that are minorities, are disproportionately concentrated in low-opportunity neighborhoods despite the program's intended goal of facilitating geographic mobility (Pendall 2000; Basolo and Nguyen 2005; Lens 2013). Recent research has illuminated the constrained housing search processes of HCV holders. For instance, landlords have approached would-be HCV holders—standing in line to collect their vouchers from the housing authority—to offer these carless prospective tenants rides to apartment showings in distressed neighborhoods (Rosen 2014). The extent to which Craigslist can make below-FMR and thus eligible units legible to HCV holders could help advance the original intent of the HCV program as a choice- and mobility-maximizing policy for low-income tenants (Newman and Schnare 1997; Schwartz et al. 2017; Boeing 2019; Boeing et al. 2020). Spot market data could also help validate the all-important FMR levels or even improve their accuracy and timeliness.

Another heavily-researched policy problem is the declining rate of long-distance mobility to high-wage MSAs (cf. Manduca 2019). Americans relocate at lower rates than they once did. From the 1950s to 2009, the proportion of households that changed residence dropped by 40%, with a steady decline since 1983 (Frey 2009). Reduced economic mobility bears broad consequences. Ganong and Shoag (2017) calculate that between 1980-2010, per-capita incomes between US states converged at only half the rate they had over the prior 100 years. For most of the 20[th] century, Americans migrated from low-income to high-income places, but in 1980 this trend began to reverse before turning sharply negative from 2005-2010. Today, high housing costs in high-wage regions reduce economic mobility and thus overall economic growth. This threatens the entire economy as housing costs stifle labor moving to where it is demanded. Herkenhoff et al. (2017) and Hsieh and Moretti (2019) attribute significant reductions in US GDP to this phenomenon.

Studies such as Ganong and Shoag's generally rely on home sale prices for the reasons enumerated earlier. But most movers seek rental housing: according to the 2016 ACS, 71% of US households that had moved since the beginning of 2015 were renters. While people looking to relocate locally may in some cases utilize soft ties, personal knowledge, and in-person visits to scout suitable housing units, people relocating from out of town are more dependent on online information exchanges to search for housing remotely.

Meanwhile, the stakes of an optimal rental housing search outcome, whether long-distance or local, are increasing. Rent burdens (tenants' rents as a share of their income) have risen across the board, even in moderate- and low-priced markets with high vacancy rates (Desmond 2018). This has in part precipitated a sharp uptick in evictions, with profound consequences for the stability of low-income families (Desmond and Perkins 2016; Desmond and Wilmers 2019). Desmond (2018) argues that research and policymaking have not caught up to these realities. Incomplete data on the housing market constraints faced by renters, especially HCV holders and low-income households facing frequent moves, are a key part of this problem. Rigorous analyses of intra- or inter-metropolitan residential mobility must consider both the home buying and rental spot markets, but our collective knowledge of the latter routinely relies on transacted-rents data as a weak proxy.

## 3. Methods

Given these problems—for both long-distance and within-metro moves, and particularly the most vulnerable lower-income renters and HCV holders—this study poses the question: to what extent do widely-available official estimates of rent misrepresent the contemporaneous rental spot market? We assess the limitations of existing rental data and the strengths and weaknesses of using spot market data such as online rental listings as a supplement.

### 3.1. Data

We analyze four data sources: 1) 2014 one-year ACS median contract rent estimates; 2) 2015 AHS recent-mover median rent estimates for households that moved between 2010-2015; 3) 2014 HUD FMRs; and 4) we adopt Boeing and Waddell's (2017) data set of rental listings posted on Craigslist in 2014. Those authors detail how the Craigslist data were collected and processed, but we briefly summarize their methods here. They developed a web scraper that ran nightly between May and July 2014, collecting every US rental listing posted in the previous 24 hours. After these data were collected, they were processed and cleaned in several steps to produce a data set more appropriate for housing market analysis. Duplicate listings and reposts were removed and the rest were filtered to discard extreme outliers, obvious spam, and typographical errors. This yields a final data set of 3 million listings which we adopt for our study.

We use the ACS/AHS's reported margins of error (MOE) to construct 90% confidence intervals (CI) around their sample-derived rent estimates in the 25 MSAs covered by the 2015 AHS. We also use ACS data to compute the economic rental vacancy rate (i.e., the share of rental housing units reported as currently available for rent). Finally, from AHS data, we calculate the subsidized/controlled share of units: i.e., either subsidized (public housing, voucher-subsidized, or privately-owned subsidized housing) or rent-controlled. The AHS is the only one of these aggregated federal government sources that allows for differentiation between rental units that are subsidized/controlled and those that are not. Unless otherwise stated, our reported AHS rents include all units regardless of subsidy or control.

### 3.2. Analysis

We measure the disparity between Craigslist asking rents and ACS rents reported for the corresponding MSA as the "ACS disparity," $\delta_{ACS}$:

$$\delta_{ACS} = \frac{\text{Craigslist median rent}}{\text{ACS median rent}} - 1 \qquad (1)$$

Next, we calculate the "AHS disparity," $\delta_{AHS}$, between Craigslist asking rents and AHS recent-mover rents for the corresponding MSA:

$$\delta_{AHS} = \frac{\text{Craigslist median rent}}{\text{AHS recent-mover median rent}} - 1 \qquad (2)$$

Then we calculate a similar "FMR disparity," $\delta_{FMR}$, to quantify discrepancies between asking rents in the most widely-used rental housing information exchange and the official rent levels that govern HCV eligibility:

$$\delta_{FMR} = \frac{\text{Craigslist } n^{\text{th}} \text{ percentile rent}}{\text{FMR at } n^{\text{th}} \text{ percentile}} - 1 \qquad (3)$$

In most markets, HUD calculates FMRs at the $40^{\text{th}}$ percentile rent from ACS data (inflation-adjusted, lagged, and with a recent-mover adjustment) to offer HCV holders access to units of slightly below-typical quality, but HUD uses the 50th percentile instead in some MSAs, including the following in this study: Houston, Philadelphia, Phoenix, and Riverside. We calculate $\delta_{FMR}$ using the $40^{\text{th}}$ or $50^{\text{th}}$ percentile rent in each MSA's Craigslist distribution, depending on which HUD uses there. HUD publishes FMRs by bedroom count, so we separately calculate $\delta_{FMR}$ for units in four categories (1, 2, 3, and 4+ bedrooms) within each MSA. Studio units are omitted from the analysis due to insufficient and inconsistent data coverage, and we omit Dallas from this analysis because its FMRs were published by ZIP code rather than for the MSA.

These three indicators ($\delta_{ACS}$, $\delta_{AHS}$, and $\delta_{FMR}$) elucidate how rent information derived from conventional governmental data sources over- or under-states contemporaneous asking rents and, particularly, the market reality faced by those searching for housing information online.

## 4. Findings

In 24 of the 25 MSAs studied, $\delta_{ACS}$ is positive: the MSA's typical asking rent exceeds the median rent reported by the ACS (see Table 2). Phoenix is the only exception. This finding conforms to expectations, given that long-term tenancy, rent control, and housing subsidies likely all hold transacted rents lower than spot market rents (to varying degrees across MSAs) during a period of economic growth. However, the roster of MSAs with an extreme disparity is noteworthy. In New York, Boston, Chicago, San Francisco, and Philadelphia—a group that includes three of the five highest-rent MSAs—the disparity is +63% or greater. In these hot markets, ACS estimates are particularly disconnected from asking rents. The ACS rent estimates' MOEs are generally small: no MSA has an MOE exceeding $16 and they negligibly impact the interpretation of $\delta_{ACS}$.

Contract rents provide the most relevant comparison to Craigslist asking rents, but the ACS only reports recent-mover rents for gross rents, which include utility costs. While this ACS limitation underscores the importance of alternative data sources like Craigslist, as a robustness check we additionally calculate an alternative, $\delta_{ACS}'$, using recent-mover gross rents (Appendix 1), and still find consistent disparities in some of the tightest rental markets even though it compares rents that include utilities (ACS) with rents that do not (Craigslist). This alternative indicator is unsurprisingly lower across the board than $\delta_{ACS}$, but both its mean and median are positive and its values remain high in expensive cities like New York and San Francisco. It also strongly correlates with $\delta_{ACS}$ (Pearson's $r$=0.99, $p$<0.01). Every data source is imperfect, but this robustness check underscores how Craigslist offers a complementary though imperfect spot market lens to the current constellation of flawed data sources.

Table 3 summarizes $\delta_{FMR}$ by MSA and bedroom count. Its general pattern remains positive, though less so than for $\delta_{ACS}$. When $\delta_{FMR}$ is averaged across bedroom counts, it is positive in 15 of the 24 MSAs. We calculate a Pearson's $r$ of -0.48 ($p$<0.05) between $\delta_{FMR}$ (averaged across bedroom counts) and the MSAs' economic rental vacancy rates. That is, the tighter the housing market, the more the asking rents exceed the HCV thresholds. These results suggest that where HCVs are most needed, their users' ability to find rental units in which they can use them is most constrained—consistent with the theory that demand-side supplements function best when sufficient supply-side opportunities exist.

Table 4 summarizes the $\delta_{AHS}$ findings. In almost all cases, the entire 90% CI of the disparity between Craigslist rents and AHS recent-mover rents lies above zero. In other words, across most MSAs and bedroom types, median Craigslist rents consistently exceed recent-mover AHS rents even when accounting for sampling error in the latter. The only exceptions are 1-bedroom units in Phoenix, and 3 and 4+ bedroom units in Seattle. We can possibly attribute some of this disparity to the weak economy and low rents on offer at the beginning of the 2010-2015 period. But this alone cannot explain the full effect: in most cases the $\delta_{AHS}$ averaged across bedrooms (Table 4) exceeds the ACS-reported contract rent growth from the equivalent period, particularly in the MSAs with the greatest overall disparities, such as New York and San Francisco. Sampling biases could also exist: if Craigslist over-represents more-expensive units or neighborhoods, we might see higher median asking rents than actually occur across the full market.

Table 5 shows how several variables, including median ACS and AHS rents, correlate across the MSAs. As we might expect, in most MSAs, $\delta_{ACS}$ is greater than $\delta_{AHS}$ but they strongly correlate. One might hypothesize that both disparities would be higher in MSAs with high recent rent growth as tenant competition boosts the asking rents of

**Table 2**. ACS disparity for the 25 MSAs (with MOE)

| MSA | $\delta_{ACS}$ |
|---|---|
| Atlanta | 6% ± 1% |
| Boston | 114% ± 2% |
| Chicago | 78% ± 1% |
| Cincinnati | 22% ± 1% |
| Cleveland | 24% ± 1% |
| Dallas | 40% ± 1% |
| Denver | 30% ± 1% |
| Detroit | 16% ± 1% |
| Houston | 24% ± 1% |
| Kansas City | 12% ± 2% |
| Los Angeles | 52% ± 1% |
| Memphis | 25% ± 3% |
| Miami | 53% ± 2% |
| Milwaukee | 28% ± 1% |
| New Orleans | 32% ± 3% |
| New York City | 120% ± 1% |
| Philadelphia | 67% ± 1% |
| Phoenix | -1% ± 1% |
| Pittsburgh | 54% ± 2% |
| Portland | 26% ± 1% |
| Raleigh | 14% ± 2% |
| Riverside | 27% ± 2% |
| San Francisco | 63% ± 2% |
| Seattle | 25% ± 2% |
| Washington | 21% ± 1% |
| **Mean** | 39% |
| **Median** | 27% |

**Table 3**. FMR rent disparity, by number of bedrooms, for the 25 MSAs

| Region | $\delta_{FMR}$ (1 BR) | $\delta_{FMR}$ (2 BR) | $\delta_{FMR}$ (3 BR) | $\delta_{FMR}$ (4+ BR) | $\delta_{FMR}$ (avg) | Economic rental vacancy rate |
|---|---|---|---|---|---|---|
| Atlanta | -6% | -15% | -28% | -27% | -19% | 8.4% |
| Boston | 59% | 46% | 44% | 65% | 53% | 3.3% |
| Chicago | 39% | 42% | 26% | 41% | 37% | 6.3% |
| Cincinnati | 0% | -2% | -12% | 7% | -2% | 6.3% |
| Cleveland | 1% | -5% | -13% | -4% | -5% | 6.6% |
| Dallas | - | - | - | - | - | 7.0% |
| Denver | 32% | 29% | 11% | 16% | 22% | 4.0% |
| Detroit | -7% | -11% | -24% | -10% | -13% | 6.2% |
| Houston | 24% | 3% | -11% | -1% | 4% | 7.1% |
| Kansas City | -13% | -17% | -32% | -23% | -21% | 7.5% |
| Los Angeles | 34% | 29% | -5% | -11% | 12% | 3.3% |
| Memphis | -7% | -16% | -24% | -20% | -17% | 10.6% |
| Miami | 32% | 20% | 3% | 18% | 18% | 7.0% |
| Milwaukee | 7% | 3% | 1% | 7% | 5% | 3.7% |
| New Orleans | -2% | -5% | -8% | -3% | -4% | 8.7% |
| New York | 52% | 60% | 57% | 78% | 62% | 4.2% |
| Philadelphia | 42% | 41% | 6% | 19% | 27% | 7.1% |
| Phoenix | -16% | -17% | -22% | -21% | -19% | 7.7% |
| Pittsburgh | 18% | 11% | 0% | -5% | 6% | 5.0% |
| Portland | 18% | 10% | -2% | 10% | 9% | 2.2% |
| Raleigh | 5% | 0% | -3% | -2% | 0% | 6.2% |
| Riverside | 17% | 13% | -4% | -7% | 5% | 6.4% |
| San Francisco | 23% | 17% | 3% | -7% | 9% | 2.7% |
| Seattle | 16% | -2% | -16% | -15% | -4% | 3.2% |
| Washington | 16% | 12% | -6% | -11% | 3% | 5.2% |
| **Mean** | 16% | 10% | -3% | 4% | 7% | 5.8% |
| **Median** | 17% | 7% | -4% | -3% | 4% | 6.3% |

Note: economic vacancy rate is the percentage of all rental units that are currently available for rent (per ACS). Does not include units that are rented but unoccupied. Results for Dallas are not reported because FMRs are reported at the ZIP code level there.

Table 4. AHS recent-mover disparity (with MOE), by number of bedrooms, for the 25 MSAs

| Region | $\delta_{AHS}$ (1 BR) | $\delta_{AHS}$ (2 BRs) | $\delta_{AHS}$ (3 BRs) | $\delta_{AHS}$ (4+ BRs) | $\delta_{AHS}$ (avg by BR type) | ACS rent increase 2010-15 | Subsidized/ controlled |
|---|---|---|---|---|---|---|---|
| Atlanta | 9% ± 6% | 3% ± 3% | 0% ± 1% | 5% ± 15% | 4% | 11% | 8% |
| Boston | 94% ± 26% | 86% ± 11% | 104% ± 30% | 58% ± 22% | 86% | 12% | 18% |
| Chicago | 59% ± 17% | 72% ± 3% | 63% ± 24% | 103% ± 49% | 74% | 10% | 12% |
| Cincinnati | 20% ± 3% | 9% ± 6% | 17% ± 11% | 61% ± 24% | 27% | 10% | 14% |
| Cleveland | 16% ± 7% | 19% ± 4% | 26% ± 7% | 40% ± 17% | 25% | 6% | 15% |
| Dallas | 32% ± 5% | 66% ± 5% | 33% ± 13% | 20% ± 7% | 38% | 17% | 6% |
| Denver | 12% ± 9% | 25% ± 9% | 23% ± 8% | 26% ± 9% | 21% | 27% | 8% |
| Detroit | 4% ± 3% | 9% ± 6% | 13% ± 12% | 50% ± 12% | 19% | 9% | 12% |
| Houston | 33% ± 6% | 19% ± 3% | 20% ± 12% | 12% ± 12% | 21% | 16% | 5% |
| Kansas City | 0% ± 7% | 8% ± 5% | 14% ± 9% | 33% ± 15% | 14% | 11% | 11% |
| Los Angeles | 46% ±11% | 33% ± 2% | 21% ± 10% | 30% ± 25% | 33% | 11% | 17% |
| Memphis | 19% ± 6% | 21% ± 9% | 29% ± 6% | 24% ± 22% | 23% | 9% | 10% |
| Miami | 53% ±7% | 37% ± 6% | 20% ± 11% | 48% ± 18% | 39% | 13% | 8% |
| Milwaukee | 13% ±5% | 21% ± 4% | 36% ± 8% | 58% ± 19% | 32% | 9% | 10% |
| New Orleans | 18% ±11% | 24% ± 8% | 34% ± 11% | 51% ± 30% | 32% | 6% | 18% |
| New York | 76% ±11% | 101% ± 18% | 103% ± 26% | 199% ± 119% | 120% | 13% | 26% |
| Philadelphia | 58% ± 13% | 61% ± 10% | 52% ± 18% | 41% ± 26% | 53% | 10% | 13% |
| Phoenix | -1% ± 5% | 0% ± 2% | 14% ± 7% | 19% ± 7% | 8% | 11% | 4% |
| Pittsburgh | 43% ± 18% | 38% ± 11% | 75% ± 19% | 85% ± 86% | 60% | 15% | 16% |
| Portland | 13% ± 5% | 17% ± 4% | 14% ± 8% | 40% ± 14% | 21% | 19% | 8% |
| Raleigh | 1% ± 2% | 6% ± 3% | 27% ± 4% | 12% ± 10% | 12% | 16% | 6% |
| Riverside | 39% ± 11% | 27% ± 6% | 28% ± 8% | 22% ± 14% | 29% | 7% | 6% |
| San Francisco | 30% ± 11% | 38% ± 12% | 52% ± 17% | 63% ± 16% | 46% | 20% | 31% |
| Seattle | 22% ± 7% | 13% ± 5% | -1% ± 4% | -4% ± 8% | 8% | 20% | 8% |
| Washington | 20% ± 7% | 25% ± 3% | 26% ± 12% | 38% ± 12% | 27% | 14% | 13% |
| **Mean** | 29% | 31% | 34% | 45% | 35% | 13% | 12% |
| **Median** | 20% | 24% | 26% | 40% | 27% | 11% | 11% |

**Table 5**. Correlation matrix of indicators across all 25 MSAs

| | $\delta_{ACS}$ | $\delta_{FMR}$ | $\delta_{AHS}$ | ACS median contract rent, 2014 | Inter-quartile range of AHS contract rent, non-HUD units, 2015 | ACS rent change 2010-15 | Subsidized/ controlled | Economic rental vacancy rate |
|---|---|---|---|---|---|---|---|---|
| $\delta_{FMR}$ | 0.92*** | | | | | | | |
| $\delta_{AHS}$ | 0.98*** | 0.90*** | | | | | | |
| ACS median contract rent, 2014 | 0.44*** | 0.47** | 0.35 | | | | | |
| Interquartile range of AHS contract rent, non-HUD units, 2015 | 0.57*** | 0.55*** | 0.48** | 0.89*** | | | | |
| ACS rent change 2010-15 | 0.05 | 0.07 | -0.12 | 0.37* | 0.44** | | | |
| Pct subsidized/controlled | 0.63*** | 0.43** | 0.57*** | 0.45** | 0.52*** | -0.01 | | |
| Economic rental vacancy rate | -0.38* | -0.48** | -0.26 | -0.56*** | -0.61*** | -0.63*** | -0.36* | |
| Pct movers from outside county | 0.11 | 0.21 | 0.01 | 0.39* | 0.43** | 0.62*** | 0.17 | -0.33 |

Note: $\delta_{FMR}$ averaged across all bedroom types. Significance noted as *$p<0.1$, **$p<0.05$, ***$p<0.01$.

increasingly scarce units. However, the correlations between rent growth from 2010-2015 and the disparity indicators are insignificant: the extent to which current census estimates understate asking rents tends to be unrelated to the rate at which metropolitan rents are rising.

MSAs with little subsidized/controlled housing (e.g., Denver, Seattle, and Portland) experienced the highest rent growth rates between 2010-2015. Conversely, some of the MSAs with the most subsidized/controlled housing experienced middling rent growth (e.g., Boston, New York, and Los Angeles; San Francisco is an exception). In the former, a greater proportion of existing tenants would have faced rent increases upon lease renewal than in the latter. This finding suggests that disparity magnitudes may be driven more by the presence of rent-restricted housing than by cyclical "hotness" (or lack thereof) in the market. In other words, census data most understate asking rents in MSAs with the most subsidized/controlled housing—the same places where housing markets are tightest, overall rents highest (per the correlation between ACS rent and $δ_{ACS}$), vacancy rates lowest, and new housing supply most constrained.

An MSA's share of subsidized/controlled rental stock correlates significantly and moderately strongly with $δ_{ACS}$ ($r=0.63$, $p<0.01$), $δ_{FMR}$ ($r=0.43$, $p<0.05$), and $δ_{AHS}$ ($r=0.57$, $p<0.01$). This makes sense theoretically. In both tight and soft housing markets, long waiting lists exist for newly available subsidized units and their landlords face little obstacle finding enough tenants to fill vacancies. Their primary concern instead may be protecting themselves from Fair Housing Act liability by advertising available units in traditional media that target vulnerable populations who have historically experienced discrimination (i.e., not Craigslist). Thus, if Craigslist listings comprise mostly unsubsidized rental units, these findings suggest that the more controlled units that exist within an MSA, the more the conventional rent data (ACS and AHS) will understate current market-rate asking rents. Moreover, if transacted-rents data include rent-controlled units held below market rates, we would similarly expect asking rents on the market to be higher.

## 5. Discussion

Most governmental and proprietary data on rents report the prevailing terms of transacted leases. This information helps planners and policymakers assess affordability, calculate FMRs, and understand what existing renters pay in relation to local income and other socioeconomic traits. However, these data do not describe the spot market in which prospective renters and landlords produce an evolving equilibrium of new rents. Asking rents could diverge substantially from data reporting transacted rents due to publication lags, market lags, subsidies, and rent control. Rent data describing the full transacted market would particularly fail to describe current market conditions in locations with rapidly rising rents or large shares of rent-controlled units. Yet it is these very cities and neighborhoods that often most concern housing scholars, practitioners, and advocates of housing justice and inclusivity.

This returns us to our original question: how well do widely-available rental data sources describe the spot market? We find that in most MSAs—and particularly those with booming job markets and constrained housing production—asking rents substantially exceed ACS, FMR, and AHS estimates. However, these findings could partly represent sampling biases: if Craigslist over-represents high-rent neighborhoods and rental units, we would expect its typical asking rents to exceed more-representative estimates. However, Craigslist does not exclusively represent the high-end rental market: about half of its listings' asking rents would be affordable for households just below the US median income and about a quarter would be affordable for households just below the corresponding median for African-American households (Boeing 2019). Craigslist hosts millions of rental listings each month, many of which are within the reach of low- and moderate-income families.

Online information exchanges have important ramifications for two groups in particular that deserve further research. First, households using HCVs to relocate to higher-opportunity neighborhoods would benefit from Craigslist's broadcasting of information if it could help them find HCV-eligible units in unfamiliar neighborhoods. Online information exchanges thus play a potentially crucial role in ongoing residential sorting and segregation (Krysan and Crowder 2017; Boeing et al. 2020). Their data exhaust could also help policymakers set FMRs to better reflect current market conditions. Planners working in hot markets might use Craigslist data (instead of conducting expensive, time-consuming surveys by hand) to petition HUD for higher FMRs—for instance, by requesting that HUD use the $50^{th}$ percentile standard, rather than the typical $40^{th}$, as has happened in several MSAs. HUD itself could even incorporate spot market data sources such as Craigslist into its FMR calculations—especially considering how these online platforms' information-broadcasting benefits could help make choice neighborhoods' affordable units more legible to disadvantaged homeseekers searching by unit characteristics to overcome their own locally-constrained market knowledge (Boeing et al. 2020).

Second, while people looking to relocate locally may have the option to utilize soft ties, tacit regional knowledge, and in-person scouting trips to identify suitable housing units, people looking to relocate from out of town depend much more on online information exchanges and rental listings to conduct housing searches. They thus play an understudied role in inter-metropolitan residential and economic mobility. On average across these MSAs, 62% of movers remained in the same county, but 20% of movers were from out of state or abroad. In cities like Boston, Washington, Portland, and Seattle, over 25% of movers were from different states or abroad (details in Appendix 2). This underscores the importance of online information exchanges for out-of-town movers in many markets.

In sum, this study's findings question planners' current ability to assess affordability barriers to residential mobility when conventional data sources substantially understate current rental market conditions. Online listings data can provide valuable insights into rapidly-evolving rental housing markets that widely-available governmental sources cannot.

However, like other data sources, Craigslist data offer both advantages and limitations. On one hand, they are timelier, include unit characteristics, and represent small-area rental markets searched by current homeseekers. On the other hand, they do not represent final transacted rents and do exhibit some sampling biases. However, online listings data do faithfully represent the *online* housing market, and, according to the most recent AHS, more renters in urbanized areas found their current homes through online listing sites like Craigslist than through any other information source. Planners must weigh these trade-offs when considering data sources to understand rental markets from multiple perspectives. A publicly available spot market data source could supplement the current constellation of data sources and narrow the information gap between owner-occupied and rental housing and between transacted rents and asking rents. This is critical given the ascending importance of the US rental market alongside dwindling affordability and economic mobility.

How can researchers and practitioners gain broader access to online rental listings data? Several different paths forward are possible. Ideally, the Census Bureau would partner with online listing sites such as Craigslist, Zillow, and RentJungle. Rental listings could then underlie a constantly updated spot market data product with the federal government's imprimatur of legitimacy and accuracy. As with all other census data, various steps would need to be taken to appropriately safeguard anonymity and privacy. Ideally, however, listings would be made available in disaggregate form with due protections, as is currently the case with PUMS and AHS microdata. Other possibilities also exist. A consortium of university-based research centers might collaborate with online listing sites to routinely acquire rental listings, compile them, and operate a data clearinghouse. Regardless of the exact model, a rich set of spot market data must be widely and publicly available for researchers and practitioners to address today's most pressing rental housing questions.

## 6. Conclusion

This study analyzed millions of Craigslist rental listings to investigate how asking rents on the spot market deviate from current governmental data sources. The ACS, AHS, and FMR are the federal government's three most widely-used aggregated data sources on rents and each has its own strengths and limitations. By looking across all three of them, we found that these conventional data sources on which planners rely typically understate contemporaneous asking rents in most MSAs. Craigslist listings (and related spot market data sources) can serve as a useful though imperfect complement to today's constellation of imperfect rental market data. Each piece of this constellation has its own unique flaws, but in concert they provide a richer nuanced view of the market from different perspectives and with different trade-offs.

This is important for planners in three key ways. First, these technology platforms are rapidly changing the housing market (cf. Fields and Rogers 2019; Porter et al. 2019) and their data exhaust offers planners a new lens to understand market behavior in what has become the dominant housing information exchange. Second, spot market data provide a particularly useful complement to transacted data in markets with rapidly rising rents, due to the importance of timeliness for planners and policymakers trying to understand local affordability and craft housing interventions. Third, HCV holders and inter-metropolitan movers in particular could benefit from online information exchanges (in very different ways) that must be considered by policymakers and practitioners as well as scholars of residential and economic mobility.

## Acknowledgments

The authors wish to thank Paul Waddell, who graciously allowed the reuse of data originally collected in his lab at UC Berkeley.

**Appendix 1**. ACS recent-mover gross rent disparity for the 25 MSAs (with MOE).

| MSA | Recent-mover gross rent ACS disparity |
|---|---|
| Atlanta | -15.4% ± 0.9% |
| Boston | 74.2% ± 2.4% |
| Chicago | 46.8% ± 1.5% |
| Cincinnati | -2.2% ± 1.5% |
| Cleveland | 0.9% ± 1.4% |
| Dallas | 14.3% ± 1.0% |
| Denver | 11.4% ± 1.5% |
| Detroit | -9.3% ± 1.0% |
| Houston | 0.5% ± 0.9% |
| Kansas City | -14.2% ± 1.3% |
| Los Angeles | 29.9% ± 0.7% |
| Memphis | -8.3% ± 1.8% |
| Miami | 31.0% ± 1.2% |
| Milwaukee | 5.4% ± 1.2% |
| New Orleans | 9.2% ± 1.8% |
| New York City | 78.4% ± 1.2% |
| Philadelphia | 34.5% ± 1.4% |
| Phoenix | -17.6% ± 1.1% |
| Pittsburgh | 17.8% ± 1.8% |
| Portland | 5.4% ± 1.6% |
| Raleigh | -7.1% ± 2.1% |
| Riverside | 7.1% ± 1.3% |
| San Francisco | 40.4% ± 1.8% |
| Seattle | 5.2% ± 1.1% |
| Washington | 5.4% ± 1.0% |
| **Mean** | 13.7% |
| **Median** | 5.4% |

**Appendix 2**. ACS movers' origins for the 25 MSAs.

| MSA | Local (Same County) Mover | Non-Local Same-State Mover | Out of State or Abroad Mover |
|---|---|---|---|
| Atlanta | 49% | 30% | 21% |
| Boston | 52% | 20% | 28% |
| Chicago | 69% | 12% | 18% |
| Cincinnati | 64% | 18% | 18% |
| Cleveland | 76% | 12% | 12% |
| Dallas | 62% | 21% | 17% |
| Denver | 44% | 32% | 24% |
| Detroit | 67% | 20% | 13% |
| Houston | 65% | 16% | 20% |
| Kansas City | 60% | 16% | 24% |
| Los Angeles | 76% | 10% | 15% |
| Memphis | 74% | 7% | 19% |
| Miami | 71% | 10% | 19% |
| Milwaukee | 70% | 17% | 13% |
| New Orleans | 59% | 21% | 20% |
| New York City | 60% | 17% | 24% |
| Philadelphia | 58% | 17% | 24% |
| Phoenix | 71% | 7% | 22% |
| Pittsburgh | 62% | 17% | 20% |
| Portland | 54% | 20% | 27% |
| Raleigh | 61% | 18% | 22% |
| Riverside | 67% | 22% | 10% |
| San Francisco | 51% | 27% | 22% |
| Seattle | 61% | 13% | 26% |
| Washington | 50% | 16% | 34% |
| **Mean** | 62.1% | 17.4% | 20.5% |
| **Median** | 62.0% | 18.0% | 20.0% |